\documentstyle[11pt,fullpage,amssymb,rotate]{article}

\input{psfig}
\begin{document}

\begin{titlepage}
\begin{center}
{\large
XIX INTERNATIONAL SYMPOSIUM ON LEPTON AND PHOTON INTERACTION AT HIGH
ENERGY \\

Stanford University, August 9-14, 1999} \\[6cm]

{\Large L.G.Landsberg\\
State Research Center, Institute for High Energy Physics,\\ 
Protvino, Moscow region, Russia, 142284} \\[3cm]

{\Large \bf
STUDY OF THE PRIMAKOFF REACTIONS \\
AT HIGH ENERGY
}
\end{center}

\end{titlepage}

\newpage
\pagestyle{plain}
\setcounter{page}{1}

\vspace*{2cm}

\begin{abstract}
In this talk the Coulomb production reactions at high energy are discussed
as well as the study of electromagnetic properties of hadrons (mesons and 
hyperons) in these processes. The results of previous investigations are 
summed together with some recent data of SELEX(Fermilab) and SPHINX(IHEP) 
experiments.
\end{abstract}

\section{Introduction}
Investigation of electromagnetic hadron decays plays an important role 
in elementary-particle physics. These processes, governed by the 
interaction of real and virtual photons with the electric charges of 
quark fields, make it possible to obtain unique information on the
character of various quark configurations in hadrons,
the mechanism  of mixing,  the electromagnetic structure of strongly
interacting particles, and on some phenomenological features of
such particles (magnetic moments, formfactors, polarizability, etc.). 
Recent years have seen a significant progress in this field.

Radiative hadron decays of the type $a \to h + \gamma$ can be studied by 
both direct and indirect methods. In the former case, particle $a$ is 
produced in a  reaction  and its  decay $a \to h + \gamma$
is detected directly in the experimental setup. Among indirect methods 
one should  mention the process of   interaction of primary particle $h$
with a virtual photon in the Coulomb field of the nucleus
\begin{equation}
h + (Z,A) \to a + (Z,A)
\end{equation}
As we will show later, at a very high energy $E_h$  Coulomb reaction (1)
is characterized by small values of the squared 4-momentum transfer
$q^2$ which is identical with the squared 4-momentum $q^2$ of the virtual
photon propagator. Thus, at $q^2\simeq 0$  the virtual photons
are quasi-real. In this approximation the
cross section for this Coulomb production process is proportional 
to the radiative width $\Gamma(a \to h + \gamma)$. 

The main aim of my talk is to discuss these electromagnetic
Coulomb processes and their application to the study of radiative 
decays of hadrons and to the search for some exotic states. Actually, 
the type (1) reactions involving collisions with virtual (quasireal)
photons are photoproduction reactions on primary
hadrons $h$. With pion, kaon, or hyperon beams we have  unique
possibility to study photoproduction on unstable ``targets" ($\pi,~K,~Y$).
The Coulomb production mechanism was considered for the first time 
independently by H.Primakoff~[1] and by I.~Pomeranchuk and
I.~Schmushkevich~[2]. It is often designated as the Primakoff production
or Primakoff reaction.

\section{Direct~Methods~of~Studying the~Radiative~Decays~of~Hadrons}

In this type of experiments the particles under study ($a$) are produced 
either in hadron reactions or in electromagnetic interactions 
(photoproduction; resonance production in $e^+e^-$ collisions). 
As a rule, to select rare electromagnetic decays
\begin{equation}
a\to h + \gamma ,
\end{equation}
\begin{equation}
a\to b_1 + b_2 +\gamma ,
\end{equation}
it is necessary to detect all the decay products --- charge particles as well
as photons --- to measure their momenta and energy and to reconstruct the
effective mass of decay particles $(M(h\gamma);~M(b_1b_2\gamma))$.

In the directly detected $a \to h + \gamma$ decays or more intricate
electromagnetic processes, we must suppress a background from the decay
$\pi^0 \to \gamma \gamma $ with one photon being lost. A similar situation prevails,
for  example, in searches for the decay $\omega \to \pi^+ \pi^- \gamma$ 
against the background from the main decay mode  
$\omega \to \pi^+ \pi^- \pi^0$, $\pi^0 \to \gamma (\gamma)$ 
[hereafter, ($\gamma$) stands for the lost photon].
Reliable separation of rare radiative decays involving single 
photons requires a complete detection of all secondaries (both charged and
 neutral ones), measurement of their momenta, and reconstruction of 
effective masses with the highest possible resolution. The background from 
the lost photons can be minimized by equipping the setup with a veto system 
consisting of counters with a low threshold for photon detection and 
having the maximum possible coverage. This system also includes a 
hodoscopic $\gamma$ spectrometer detecting photons from radiative decays 
in the  operating acceptance of the setup. A strong kinematical constraint 
on the processes in question can also play an important role. The point is 
that the background from the events with a lost photon does not yield a narrow
peak at the mass of a particle under study and can therefore be considerably 
reduced owing to kinematical constraints. The above is illustrated by searches
for the radiative decay $\omega \to \pi^+ \pi^-\gamma$ in the experiments with
the Lepton-F [3] and ASTERIX [4] setups. The background from the lost 
photons was substantially suppressed by appropriately chosen conditions and
using a veto system in the former case and by imposing more stringent 
kinematical constraints in the latter case. As a result, the two experiments 
yield approximately the same upper limit on the probability of the above 
radiative decay: $BR(\omega \to \pi^+ \pi^- \gamma)<4 \times 10^{-3}$ 
(at a 95\% C.L.) This demonstrates that 
radiative decays can be sought directly even if their 
probabilities are less than 1\% of the probabilities of the most 
dangerous background processes. At the same time, the decay 
$\omega \to \pi^0 \pi^0 \gamma$, for which the background conditions proved 
to be much more favorable (the process $\omega \to 3\pi^0$, which could
mimic the decay $\omega \to \pi^0 \pi^0  \gamma$ because of the lost 
photon, is forbidden by $P$ and $C$ conservation in strong and electromagnetic
interactions), was recorded at a significantly lower level of 
$BR(\omega \to \pi^0 \pi^0 \gamma)=(7.2 \pm 2.6) \times 10^{-5}$~[5].
The same is true for several other electromagnetic decays with a relatively
small background. The results of several experiments for the direct study
of rare radiative decays of mesons are presented in Table 1 and in 
[3-10]. An overview of various methods for 
detecting electromagnetic decays is given, for example, in review papers 
[11-14].


\section{Coulomb production of hadrons and the study of their radiative 
decays}
To determine the radiative widths of some excited hadrons for which direct 
measurements involve considerable difficulties or are impossible, 
it was proposed long ago to use coherent electromagnetic production of 
particles in the Coulomb field of heavy nuclei ([1,2]),
see also [15-18]). 
In the small-width approximation as 
applied to the resonance state $a$ in (1), the cross section 
for electromagnetic production has the from
\begin{equation}
\begin{array}{c}
\displaystyle{
\nonumber \left[\frac{d\sigma[h+(Z,A)\to a + (Z,A)]}{dq^2} \right]_{\rm Coul}=}\\
\displaystyle{
=|T_{\rm Coul}|^2=8\pi \alpha Z^2 
\cdot \frac{2J_a+1}{2J_h+1} \cdot \Gamma(a \to h \gamma)}\\
\displaystyle{
\nonumber\times \left[\frac{(q^2 - q^2_{\min})}{q^4} \right] \cdot 
\left(\frac{M_a}{(M^2_a-M^2_h)^3}\right)^3 \cdot 
|F_Z(q^2)|^2,}
\end{array}
\end{equation}
where $Z$ is the charge  of the nucleus; $\alpha$ is the fine-structure
constant; $\Gamma(a \to h\gamma)$ is the 
radiative decay  width of particle $a$; $J_a$ and $J_h$ are the 
spins of particles $a$ and $h$, respectively; $M_a$ and $M_h$ are their masses;
$F_Z(q^2)$ is the nuclear electromagnetic form factor; 
$q^2_{\min}=(M_a^2-M_h^2)^2/4E^2_h$ is the minimum 4-momentum transfer squared;
and $E_h$ is the energy  of the primary particle $h$. If photons appear as 
primary particles, an additional factor of two must be included in the 
expression for the differential cross section.

As it is seen from (4)
the Coulomb cross section increases fast with decreasing 
$q^2$. It can easily be found that the cross section 
$[d\sigma/dq^2]_{\rm Coul}$ attains a maximum at $q^2_0=2q^2_{\min}$
and that $[d\sigma/dq^2]_{\rm Coul}^{\max} \propto q^{-2}_{\min}
\propto E^{2}_h$. As the primary momentum increases, the cross-section 
value at the  maximum increases in proportion to $E^2_h$, whereas the 
position of this maximum shifts toward lower values of $q^2$. 
The total cross section for the Coulomb process  increases
as $\ln E_h$. Concurrently,
the differential  cross section for a coherent process that is governed by 
a strong interaction exhibits a much broader distribution in $q^2$, and the
region of small $q^2$ is dominated by the Coulomb contribution.
The total Coulomb production cross section has the form
\begin{equation}
\begin{array}{c}
\sigma[h+(Z,A)\to a + (Z,A)]_{\rm Coul}=\\
\displaystyle{
=\frac{2J_a+1}{2J_h+1}\cdot
8\pi \alpha Z^2 \Gamma(a \to h \gamma) 
\left(\frac{M_a}{M^2_a-M^2_h}\right)^3} \times \\
\displaystyle{
\nonumber \times \int^{q^2_{\max}}_{q^2_{\min}}
\frac{(q^2 - q^2_{\min})}{q^4}|F_Z(q^2)|^2 d(q^2)=
\sigma_0 \cdot \Gamma(a \to h \gamma).
}
\end{array}
\end{equation}

The quantity $q^2_{\max}$ bounds the integration domain dominated by
Coulomb processes. Certainly, a more precise expression for $\sigma_0$
can be obtained and used in the data analysis, in which 
$a\to h \gamma$ decay is expressed more carefully, in a relativistic 
Breit-Wigner resonance form.

The coherent production of hadrons by a primary particle $h$ on nuclei 
$(Z,A)$ is a rather complicated process affected by both electromagnetic
and strong interactions. The contribution of strong interactions leads to a
broader distribution in momentum transfer squared $q^2$. The differential
cross section for coherent particle production on nuclei can be represented as
\begin{equation}
\left[d\sigma/dq^2\right]_{\rm Coherent}=\left|T_{\rm Coul.} + e^{i\varphi}
T_{\rm strong}\right|^2,
\end{equation}
where $T_{\rm Coul.}$ is the amplitude of Coulomb production, 
$T_{\rm strong}$ is the amplitude of a coherent process that is governed by
a strong interaction, and $\varphi$ is the  relative phase of the
two amplitudes. For the understanding of a possibility to separate these
coherent processes it is instructive to formulate their main characteristics
which is done in Table 2.

It can be seen from Table 2 that in accordance with quantum 
numbers of the produced hadrons in (1) two possible  scenarios for
the strong coherent background can take place:
\begin{enumerate}
\item[A.] Coherent diffractive background is forbidden by selection rules for
(1) and the leading coherent background is determine usually by $\omega$
reggeon exchange in t channel of the reaction. This amplitude  dies at a very
high energy. In several cases this background is also reduced in the region of 
small $q^2$ (as $\propto q^2-q^2_{\min}$). The absence of diffractive 
coherent background significantly simplifies the separation 
of the Coulomb process. 
\item[B.] Coherent diffractive background plays a leading role in reaction (1).
In this case the separation of the Coulomb process is much more difficult and 
might be performed only in the region of very small $q^2(<10^{-3}$ Gev$^2$)
and with a very large statistics. 
\end{enumerate}
Let us complete the discussion of the Coulomb production reactions making
several conclusions:
\begin{enumerate}
\item[1.] The Coulomb production reactions open a new possibility
to study rare radiative decays $a\to h + \gamma$ if these processes 
cannot be detected directly
due to the background from the decay $a \to h + \pi^0$; 
$\pi^0 \to \gamma + (\gamma)$ with lost photons or if the decay width is too 
small for a direct measurement (the later case takes place, for example, for
$\Sigma^0  \to \Lambda + \gamma$ decays). 
\item[2.] The Coulomb production processes open a unique possibility to study the
photoproduction reactions on unstable ``targets" (like pions, kaons, hyperons),
which can be  very important for the search for exotic hadrons, for studying some
electromagnetic properties 
of unstable hadrons (their polarizability, for example)
and other phenomena.
\item[3.] To accomplish  this study it is necessary to separate the
Coulomb production processes and to make the absolute measurements of 
their cross sections. All these require very precise and sophisticated
experiments in the high energy region. This is the reason for a relatively
small amount of the Coulomb production experiments which have been carried out in a 50
year period of time since the first theoretical work of H.Primakoff [1].
The main experimental results in this field
at  the high energies  [19--31] are summarized in Table 3.
\end{enumerate}

\section{Study of the Coulomb production processes in the SELEX experiment 
at $E_h=600$ GeV}
In 1998 the first preliminary results of studying several Coulomb production 
reactions in the SELEX experiment at the Tevatron of Fermilab were obtained
and are presented here in short (for more details see [32,33]).

The SELEX setup [34,35], that is used in these measurements includes the
three-stage magnetic spectrometer with proportional and drift chambers, 
vertex microstrip detector, trigger hodoscopes, some additional microstrip
detectors, RICH and TRD detectors for particle identification, three photon 
lead glass multichannel spectrometers, neutron calorimeter. The main 
measurements in the SELEX (E781) experiment were performed with a
negative-charged beam of 600~GeV momentum  and with practically equal
amount of $\pi^-$-mesons and $\Sigma^-$-hyperons
(they were identified with the beam TRD detector). The wide research program
is performed by SELEX(E781) international collaboration (with the participation
of scientists from Brasil, China, Germany, Israel, Italy, Mexico, Russia, 
United Kingdom, USA). The main part of this program involves the study
of charmed and strange-charmed baryons,  the search for exotic states,
the investigation of electromagnetic properties of hadrons and the
Coulomb production processes, some other researches.
A short summary of the preliminary Coulomb production results in the SELEX
experiment is presented below.

\subsection{Study of the reaction $\boldmath \pi^- + (A,Z) \to [\pi^-\pi^-\pi^+]+(A,Z)$
and measurement of the
radiative width for $\boldmath a_2(1320)$ meson}

The coherent reactions 
\vspace*{-0.2cm}
\begin{equation}
\pi^- + (Z,A) \to [\pi^-\pi^-\pi^+]+(A,Z)
\vspace*{-0.2cm}
\end{equation}
on carbon, copper and lead nuclei were separated in the analysis of the events
with 3 charge particles in the final state in the interactions of 
identified primary $\pi^-$-mesons in one
of the targets. These events must satisfy the 
``elastic conditions" --- the difference between the beam particle momentum
and the sum of momenta of three
secondary particles must be less than
17.5~GeV. The statistics of the selected events of (7) is tabulated in
Table 4.

The $P^2_T$ distribution\footnote{Here $q^2 \simeq q^2_{\min} +
P^2_T \simeq P^2_T$ at a very high momentum.}
of the $(3\pi)^-$-system in reaction (7) for 
the copper target is shown, as an example, in Fig.1a. This $dN/dP^2_T$
distribution was fitted by the sum of two exponents 
$dN/dP^2_T=C_1exp(-b_1P^2_T)+C_2exp(-b_2P^2_T)$ 
with the slopes $b_1=1577 \pm 88$ GeV$^{-2}$ and 
$b_2=189.4  \pm 0.6$ GeV$^{-2}$. The second slope $b_2$  
is in 
a good agreement with the slope for diffractive production coherent 
process (7) obtained from the data [36] at $E_{\pi}=200$ GeV. The slope $b_1$ 
corresponds to the Monte Carlo estimations for the Coulomb mechanism
(with account for the $P^2_T$ resolution).


Two $P^2_T$ regions were defined for  $dN/dP^2_T$ distribution (see Fig.1a). 
In the first 
region ($P^2_T<0.001$ GeV$^2$) the Coulomb production process dominating. The second region
($0.0015<P^2_T<0.0035$ GeV$^2$) was used to estimate the coherent diffractive background. The mass
spectra $M(3\pi)$ for these two regions are presented in Fig.1b. 
The result of  diffractive background subtraction (after a proper normalization)
is presented in Fig.1c. In this subtracted mass spectrum the Coulomb
production of $a_2(1320)^-$ meson in the reaction 
\vspace*{-0.2cm}
\begin{eqnarray}
\pi^- + Cu &\to &a_2(1320)^- + Cu \\
\nonumber && \hspace{2mm}^| \hspace{-2mm} \to \pi^- \pi^- \pi^+
\vspace*{-0.8cm}
\end{eqnarray}
is clearly seen. The same results were also obtained from the measurements with
carbon and lead targets. After the absolute normalization of cross sections for
the Primakoff reactions 
$ \sigma_{\rm Coulomb}[a_2(1320)^-]=\sigma_0\cdot \Gamma [a_2(1320)^-
\to \pi^- \gamma]$, the values for radiative width
$\Gamma [a_2(1320)^- \to \pi^- \gamma]$ from all these measurements were
obtained (see Table 4). For the absolute normalization the  
cross sections of diffractive production reactions (7) 
for $A= Pb$; $Cu$; $C$ from [36] were used. 

The data for the radiative width in the measurements with different targets are
consistent with each other, which confirms the Coulomb production of
$a_2(1320)$ meson. The average value of radiative width over all the targets is
\vspace*{-0.2cm}
\begin{equation}
\Gamma[a_2(1320)^- \to \pi^- \gamma]=233 \pm 31 ({\rm stat.}) \pm
47({\rm syst.)~KeV}.
\vspace*{-0.2cm}
\end{equation}
The main source for the systematic uncertainty is the absolute normalization
procedure. This value of the radiative width is preliminary and might be 
obtained with a better accuracy in the future. The comparison of value (10)
with previous measurement [21] and with theoretical predictions [37, 38]
is also presented in Table 4.

\subsection{Further Primakoff reaction data from the SELEX experiment}
\subsubsection{\rm Study of the coherent reactions}
\vspace*{-0.4cm}
\begin{eqnarray}
\pi^- +(A,Z)  & \rightarrow & [\pi^- \pi^- \pi^+ \pi^0]+ (A,Z), \\
\nonumber && \hspace{17mm}^|\hspace{-2mm} \rightarrow \gamma \gamma
\vspace*{-0.7cm}
\end{eqnarray}
\vspace*{-0.5cm}
\begin{eqnarray}
&\rightarrow&[\pi^- \omega ] + (A,Z), \\
\nonumber && \hspace{7mm}^|\hspace{-2mm} \rightarrow \pi^+ \pi^-  \pi^0 \\
&\rightarrow & b_1(1235)  + (A,Z), \\ 
\nonumber && \hspace{2mm}^| \hspace{-2mm} \rightarrow \omega \pi^- \\
&\rightarrow & [\pi^- \eta] + (A,Z), \\
\nonumber && \hspace{6mm}^| \hspace{-2mm} \rightarrow \pi^+ \pi^- \pi^0  \\
&\rightarrow & a_2(1235) + (A,Z), \\
\nonumber && \hspace{2mm}^| \hspace{-2mm} \rightarrow  
\pi^- \eta 
\vspace*{-0.2cm}
\end{eqnarray}
were performed at $E_{\pi^-}=600$ GeV on the lead, copper and carbon targets
in the SELEX experiment. The Coulomb production of $b_1(1235)^- \to 
\omega \pi^-$ and $a_2(1320)^- \to \eta \pi^-$ was observed quite clearly 
in (12) and (14) under very favorable background conditions.
The Coherent strong interaction background here is dominated by $\omega$
exchange and is small in the region of $P^2_T<0.003$ GeV$^2$. Now the data 
from these Primakoff reactions are in the course of analysis to obtain the absolute
normalization of their cross sections and  
to determine the 
radiative widths $\Gamma[b_1(1235)^- \to \pi^- + \gamma]$ 
and $\Gamma[a_2(1320)^- \to \pi^- + \gamma].$

\subsubsection{\rm Study of the Coulomb production reaction}
\begin{equation}
\Sigma^- + Pb \to \Sigma^*(1385)^- + Pb.
\end{equation}
This Primakoff process is determined by a very small width of 
SU(3) forbidden radiative decay
\begin{equation}
\Sigma^*(1385)^- \to \Sigma^- + \gamma.
\end{equation}
Only  a short dedicated run at the SELEX facility can be used for
the study of Primakoff reaction (15). The preliminary value for the upper
limit of radiative width for decay (16) is obtained in this measurement
\begin{equation}
\Gamma[\Sigma^*(1385)^- \to \Sigma^- + \gamma]< 7 {\rm KeV}~~({\rm 95 c.l.})
\end{equation}
The theoretical expectations for this width are in the  region of 
1-- 10 KeV (see  reviews [39,40] and the references herein). For
comparison, the expected value of radiative width for the SU(3) allowed decay
\begin{equation}
\Sigma^*(1385)^+ \to \Sigma^+ + \gamma
\end{equation}
is predicted at a level of 100~keV.

\section{Coulomb reactions and the search for exotics}

The Coulomb production processes (``Primakoff reactions") can be of great
concern not only in the study of radiative 
hadronic decays, but in the search for new types of  hadrons as well.
In the last decade the problem of existence of a novel form of hadronic 
matter~--- exotic hadrons~--- has become the  leading direction in hadron
spectroscopy. A rapid development of this field  led to a significant
advance in the systematics of ``ordinary" hadrons (with the  valence
quark structure $q \bar q$ or $qqq$) and to the observation of several unusual
states which do not fit  this simplest systematics. These anomalous states
are  real candidates for exotic hadrons with a complicated valence internal
structure (multiquark formations $qq \bar q \bar q$, $qqqq \bar q$; 
hybrid states with valence quarks and gluons $q \bar q g$, $qqqg$; pure 
gluonic mesons --- glueballs $gg,ggg$).

The success of experiments aimed at the search for exotic hadrons and 
first of all for cryptoexotic states with usual quantum numbers but 
with anomalous dynamical properties (``hidden exotics") is, to a great extent,
determined by the appropriate choice of the production processes for 
which some qualitative considerations can predict more distinct  
manifestations of exotic states. For example, it was emphasized in a 
number of studies that gluon-rich diffractive production reactions on 
nucleons or nuclei (coherent reactions) offer some favorable grounds for 
exotic hadron production [41--44]. It was also stated that for some cases the 
electromagnetic mechanisms can be very promising for these aims [45--47].

Let us consider, for example, the exotic hadrons with hidden strangeness 
--- $q\bar q s \bar s$ mesons or $qqqs \bar s$ baryons 
(here $q=u$- or $d$-quarks). Due to a significant coupling of photon 
with $s \bar s $ pairs through $\phi$ meson in the framework of VDM, 
electromagnetic interactions can provide a natural way to embed the 
$s \bar s$ pair into an intermediate hadron state and to produce the
exotic hadron with hidden strangeness. This possibility
can be realized in
the resonance $s$-channel reaction
\begin{equation}
\nonumber \gamma + N \to |qqq s \bar s \rangle \to YK
\end{equation}
or in the Coulomb production reaction  
\begin{equation}
h + (Z,A) \to |qqq s \bar s \rangle + (Z,A)
\end{equation}
As has been shown in Section 3, the Coulomb production
mechanism plays the leading role in the region of very small transfer 
momenta, where it is dominating over the strong interaction process. Thus, the
coherent reactions at a very small $P^2_T(\lesssim 0.01$ GeV$^2$) can be 
used in the search for exotic states with hidden strangeness.

As was stated before, the most interesting Coulomb production processes were
those initiated by unstable primary particles ($\pi$-mesons, $K$-mesons, 
hyperons), because these processes give the only possibility to study the 
photoproduction reactions on unstable targets. Reaction (20) 
for protons $(h\equiv p)$ is, in principle, the same as photoproduction process (19)
which can be studied in a very detailed way on photon beams of electron 
accelerators. But the existing data for these reactions are rather poor 
now and insufficient for such systematical studies.  One can hope that 
in the near future reliable data would be obtained in the experiments at
strong current electron accelerators CEBAF and ELSA (see, for example, [48]).

On the other hand, there are  the data for coherent reactions
\begin{equation}
p + C \to [Y^0 K^+] + C
\end{equation}
obtained in the experiments of the SPHINX Collaboration which might be 
connected with the Coulomb production mechanism, [47,49].

The feasibility to separate the Coulomb production processes in the 
coherent proton reactions at $E_p=70$ GeV on the carbon target in the 
measurement with the SPHINX setup has been recently demonstrated in the observation
of the Coulomb production of $\Delta(1232)^+$ isobar with $I=3/2$ in the
reaction
\begin{equation}
p + C \to \Delta(1232)^+ + C
\end{equation}
(see [47]).

\section{Conclusion}
The recent results of  the Coulomb production reactions study presented in
this talk allow one to obtain the important information on the
electromagnetic properties of hadrons and to search for new hadronic
structures. We continue the analysis of the data obtained in the SELEX
experiment and we hope to improve the precision of these
measurements and the 
number of Primakoff reactions under study. 

It is a pleasure for me to express  gratitude to my colleagues from
the SELEX and SPHINX Collaborations for a tight cooperation in obtaining
the main results presented here.

The work is partially supported by RFBR (grant 96-02-16759a).

\newpage
\begin{table}[h]
\vspace*{-1cm}
\small
\caption{Direct study of rare radiative decays of light mesons.}
\label{tab-1}
\rotate[l]{ 
\begin{tabular}{|p{30mm}|p{67mm}|p{60mm}|p{45mm}|}
\hline
Experiment & Process under study  & Result & Note  \\ \hline 
LEPTON-F \newline (IHEP) [3] \newline $\omega \to \pi^+ \pi^- \gamma$ & 
$
\begin{array}[t]{l}
\pi^- + p \rightarrow \omega + n~~(P_{\pi^-}=38~{\rm GeV}) \cr 
\phantom {\pi^- + p \rightarrow } \hspace{2mm}^| \hspace{-2mm} 
\rightarrow \pi^+ \pi^- \gamma
\end{array}
$
&
$BR[\omega \rightarrow \pi^+ \pi^- \gamma]<4 \cdot 10^{-3}$
(95\% C.L.) & Complicated \newline  background conditions due to
\newline the main decay channel\\ 
 \cline{1-3}
ASTERIX \newline (CERN) [4]  \newline  $\omega \to \pi^+ \pi^- \gamma$ & 
$
\begin{array}[t]{l}
\bar p p \rightarrow \pi^+ \pi^- \omega~~~~~~~~~~~~~~~({\rm annihilation} \cr
\phantom{\bar p p \rightarrow \pi^+ \pi^-} \hspace{1mm}^| \hspace{-2mm} 
\rightarrow \pi^+ \pi^- \gamma ~~~~ {\rm at~ rest})
\end{array}
$
&
$BR[\omega \rightarrow \pi^+ \pi^- \gamma]<4 \cdot 10^{-3}$
(95\% C.L.) & with one lost photon:  \newline 
$\omega \to \pi^+ \pi^- \pi^0 \to$  \newline 
$\to \pi^+ \pi^- \gamma (\gamma)$ \\ \hline 
GAMS-2000 
& $\pi^- + p \rightarrow [\pi^0\pi^0\gamma]+n~~(P_{\pi^-}=38$ GeV)& 
$BR[\omega \rightarrow \pi^0 \pi^0 \gamma]=$ & Favourable background \\ 
(IHEP-CERN)[5]  & 
$
\begin{array}[t]{l}
\phantom{\pi^- + p}\rightarrow \omega + n \cr 
\phantom{\pi^- + p \rightarrow}\hspace{2mm}^|\hspace{-2mm} 
\rightarrow \pi^0 \pi^0 \gamma
\end{array}
$
&
=$(7.2 \pm 2.6)\cdot 10^{-5}$ & conditions \\
$\omega \rightarrow \pi^0 \pi^0 \gamma$ & 
$40 \pm 12$ events of the decay  $\omega \rightarrow \pi^0 \pi^0 \gamma$
\newline have been observed & & 
(no decays $\omega {\times}{\hspace*{-0.4cm}}\to  3 \pi^0 \to$
\newline $\to \pi^0 \pi^0 \gamma (\gamma)$) \\ \hline
GAMS-2000 
& $\pi^- + p \rightarrow \eta +n~~(P_{\pi^-}=38$ GeV)& 
$BR[\eta  \rightarrow \pi^0 \gamma  \gamma]=$ & Favourable background \\ 
(IHEP- & 
$
\begin{array}[t]{l}
\phantom{\pi^- + p \rightarrow}\hspace{1mm}^|\hspace{-2mm} 
\rightarrow \pi^0 \gamma \gamma
\end{array}
$
&
=$(9.5 \pm 3.2)\cdot 10^{-4}$ & conditions (main background \\
CERN)[6] & Around 70 events $\eta  \rightarrow \pi^0 \gamma \gamma$
&& is with 2 lost $\gamma$: \\
$\eta \to \pi^0 \gamma \gamma$ &   
 have been observed & &
$\eta \to 3 \pi^0 
 \to \pi^0 \gamma (\gamma) \gamma (\gamma)$) \\ \hline
LEPTON-F  & 
$\pi^- + p \rightarrow K^+K^- \gamma +n~~(P_{\pi^-}=32,5$ {\rm GeV})&
$BR[f_1(1285)\rightarrow \phi \gamma]=$ & Favourable background \\
(IHEP) [7,8] & 
$
\begin{array}[t]{l}
\phantom{\pi^- + p}\rightarrow \phi  \gamma \hspace{8.5mm} +n\cr 
\phantom{\pi^- + p \rightarrow} \hspace{2mm}^|\hspace{-2mm} 
\rightarrow K^+ K^-
\end{array}
$
&
=$(0.9 \pm 0.2 \pm 0.4)\cdot 10^{-3}$ & conditions (no decays \\
$D/f_1(1285) \rightarrow \phi \gamma$ &
$19\pm 5$ events of the decay
$D/f_1(1285) \rightarrow \phi \gamma$
\newline 
have been detected& &
 $D/f_1 {\times}{\hspace*{-0.4cm}} \to \phi \pi^0 \to \varphi \gamma (\gamma))$\\
\hline
LEPTON-G & $\pi^- + p \rightarrow [\mu^+ \mu^- \gamma]+ n 
~~(P_{\pi^-}=32,5$ {\rm GeV})& 
$BR[\eta \rightarrow  \mu^+ \mu^- \gamma]=(3.4 \pm 0.4)\cdot 10^{-4}$
& Favourable background \\
(IHEP) [9-11] & 
$
\begin{array}[t]{l}
\phantom{\pi^- + p}\rightarrow \eta; \eta' \hspace{6mm} +n\cr 
\phantom{\pi^- + p \rightarrow}\hspace{2mm}^| \hspace{-2mm} 
\rightarrow \mu^+ \mu^- \gamma
\end{array}
$
& $BR[\eta' \rightarrow  \mu^+ \mu^- \gamma]=(8.9 \pm 2.4)\cdot 10^{-5}$
& conditions (no decays \\
$\eta \rightarrow  \mu^+ \mu^- \gamma$ \newline 
$\eta' \rightarrow  \mu^+ \mu^- \gamma$& 
$\sim 600$ events of the decay  
$\eta \rightarrow  \mu^+ \mu^- \gamma$ and \newline 
$33 \pm 7$ events of the decay  
$\eta' \rightarrow  \mu^+ \mu^- \gamma$ have been observed &
Measurements of electromagnetic \newline formfactors of $\eta$ and 
$\eta'$ mesons &
$\eta, \eta' \to \mu^+ \mu^- \pi^0 \to$ \newline 
$\to \mu^+ \mu^-\gamma(\gamma))$ \\
\hline 
\end{tabular} 
}
\end{table}

\newpage

\begin{table}[p]
\small
\caption{The main features of the Coulomb production processes and
the coherent background reactions governed by strong interactions 
$(h+(Z,A) \to a + (Z,A))$.}
\label{tab-1}
\vspace*{-0.5cm}
\begin{center}
\begin{tabular}{|p{50mm}|p{50mm}|p{50mm}|}
\hline 
Coulomb production processes &
\multicolumn{2}{c|}{Coherent strong interaction reactions} \\ \cline{2-3}
&$\omega$ exchange & Pomeron exchange (diffraction)
\\ \hline 
a) $\sigma_{\rm Coulomb} \propto ln E_h$. 
\newline 
b) $[d\sigma/dq^2]_{\rm Coulomb}$ is in the region of very small  $q^2$ 
(maximum of cross section is at $q^2_0=2q^2_{\min}=
\newline =2[\frac{M^2_a-M^2_h}{2E_h}]^2$. 
\newline 
c) The width of $[\frac{d\sigma}{dq^2}]_{ \rm Coulomb}$~~~ distribution is 
$\Delta \sim 6q^2_{\min}$.
\newline 
d) $\sigma_{\rm Coulomb} \propto Z^2$. 
\newline 
e) In the Gottfrid-Jackson system the $t$-channel 
\newline
helicity is \newline
 $\lambda = \pm 1$, which corresponds to quasi-real virtual photon.
&
a) $\sigma_{\rm strong~coh.} \sim E^{-1}_h$,  
e.g. this process died out at the high energies.
\newline 
b) $[d\sigma/dq^2]_{\rm strong~coh.}$ depends on
quantum numbers of particles $h$ and $a$; in some cases this cross section 
is reduced in the region of small $q^2$ as
\newline 
$[\frac{d\sigma}{dq^2}]_{\rm strong~coh.}
\propto (q^2 - q^2_{\min})$.
\newline 
c) $\sigma_{\rm strong~coh.} \propto A^{2/3}$. 
\newline 
As a rule this background is small in the region 
$q^2 \leq 0.003-0.01$ GeV$^2$. 
&
a) $\sigma_{\rm strong}$ is practically 
\newline energy independent 
\newline (Pomeron exchange). 
\newline 
b) $[\frac{d\sigma}{dq^2}]_{ \rm strong~~coh.}\propto  
\newline exp[-(q^2 - q^2_{\min})b]$
\newline
with  $b\simeq 10\cdot A^{2/3}$ GeV$^{-2}$.
\newline   
c) $\sigma_{ \rm strong~~ coh.} \propto A^{2/3}$.
\newline 
d) $t$-channel helicity  $\lambda=0$.
 \newline 
Diffractive coherent processes play a main role if they are allowed by quantum
numbers:
\newline 1)  $h$ and  $a$ 
have the same flavours;
\newline 2) $h(J^P) \rightarrow \newline
\to
a[J^P; (J+1)^{-P}; 
(J+2)^P]$ --- Gribov-Morrison selection rule. 
\\
\hline
\end{tabular}
\end{center}
\end{table}

\newpage

\begin{table}
\small
\caption{Results of the Primakoff production processes study.}
\label{tab-1}
\rotate{
\begin{tabular}{|p{30mm}|p{90mm}|p{65mm}|c|}
\hline
Experiment & Process under study & Results & Ref. \\
\hline
E272(Fermilab)& 
$
\begin{array}[t]{l}
\pi^- + (A,Z) \rightarrow \rho(770)^- +(A,Z)~~(P_{\pi^-}=156;260 {\rm GeV})\cr 
\phantom{\pi^- + (A,Z) \rightarrow}
\hspace{2mm}^| \hspace{-2mm} \rightarrow \pi^+ \pi^0
\end{array}
$
& $\Gamma[\rho(770)^- \rightarrow \pi^- + \gamma] =71 \pm 7$ keV & [20]\\
&
$
\begin{array}[t]{l}
\pi^+ + (A,Z) \rightarrow \rho(770)^+ +(A,Z)~~(P_{\pi^+}=202 {\rm GeV})\cr 
\phantom{\pi^- + (A,Z) \rightarrow}
\hspace{2mm}^| \hspace{-2mm} \rightarrow \pi^+ \pi^0
\end{array}
$
&$\Gamma[\rho(770)^+ \rightarrow \pi^+ + \gamma]=59.8 \pm 4.0$ keV & [19]\\
&
$
\begin{array}[t]{l}
\phantom{\pi^- + (A,Z) }\rightarrow a_2(1320)^+ + (A,Z) \cr
\phantom{\pi^- + (A,Z) \rightarrow} \hspace{2mm}^| \hspace{-2mm}
\rightarrow \eta \pi^+;K^0_sK^+
\end{array}
$
&
$\Gamma[a_2(1320)^+ \rightarrow \pi^+ + \gamma] =295 \pm 60$ keV & [21]\\
&
$
\begin{array}[t]{l}
\phantom{\pi^- + (A,Z) }\rightarrow a_1(1260)^+ + (A,Z) \cr
\phantom{\pi^- + (A,Z) \rightarrow} \hspace{2mm}^| \hspace{-2mm}
\rightarrow 3 \pi
\end{array}
$
&
$\Gamma[a_1(1266)^+ \rightarrow \pi^+ + \gamma] =640 \pm 246$ keV & [22]\\
&
$
\begin{array}[t]{l}
\phantom{\pi^- + (A,Z) }\rightarrow b_1(1235)^+ + (A,Z) \cr
\phantom{\pi^- + (A,Z) \rightarrow} \hspace{2mm}^| \hspace{-2mm}
\rightarrow \omega \pi^+
\end{array}
$
&
$\Gamma[b_1(1235)^+ \rightarrow \pi^+ + \gamma] =236 \pm 60$ keV & [23]\\
&
$
\begin{array}[t]{l}
K^- + (A,Z) \rightarrow K^*(890)^- + (A,Z)~~(P_{K^-}=156 {\rm GeV}) \cr
\phantom{\pi^- + (A,Z) \rightarrow} \hspace{2mm}^| \hspace{-2mm}
\rightarrow K^0_s \pi^-
\end{array}
$
&
$\Gamma[K^{*}(890)^- \rightarrow K^- + \gamma] =62 \pm 12$ keV & [24]\\
\hline
Fermilab &
$
\begin{array}[t]{l}
K^0_L + (A,Z) \rightarrow K^*(896)^0 + (A,Z)~~(P_{K^0}=60-200 {\rm GeV}) \cr
\phantom{\pi^- + (A,Z) \rightarrow} \hspace{2mm}^| \hspace{-2mm}
\rightarrow K^0_s \pi^0
\end{array}
$
&
$\Gamma[K^{*}(896)^0 \rightarrow K^0 + \gamma] =116.5 \pm 9.9$ keV & [25]\\
&
$
\begin{array}[t]{l}
\phantom{K^- + (A,Z) }\rightarrow K^*(1430)^0 + (A,Z) \cr
\phantom{\pi^- + (A,Z) \rightarrow} \hspace{2mm}^| \hspace{-2mm}
\rightarrow K^0_s \pi^0
\end{array}
$
&
$\Gamma[K^*(1420)^0 \rightarrow K^0 + \gamma] <84$ keV
\newline (90\% C.L.) & [26]\\ 
\hline
CERN& 
$
\begin{array}[t]{l}
\pi^- + (A,Z) \rightarrow \rho(770)^- +(A,Z)~~(P_{\pi^-}=200 {\rm GeV})\cr 
\phantom{\pi^- + (A,Z) \rightarrow}
\hspace{2mm}^| \hspace{-2mm} \rightarrow \pi^- \pi^0
\end{array}
$
& $\Gamma[\rho(770)^- \rightarrow \pi^- + \gamma] =81 \pm 4 \pm 4$ keV & [27]\\
\hline
SIGMA(IHEP)& 
$
\begin{array}[t]{l}
\pi^- + (A,Z) \rightarrow \pi^- \gamma +(A,Z)~~(P_{\pi^-}=40 {\rm GeV})\cr 
\end{array}
$
& 
$\beta_{\pi}=(-7.1 \pm 2.8 \pm 1.8)\cdot 10^{-43}$ cm$^3$ & [28]\\
& Compton  $\gamma \pi$-scattering; measurements of magnetic $(\beta_{\pi})$ and
electric  $(\alpha_{\pi})$ polarizability of pion &
 $\alpha_{\pi} + \beta_{\pi}=(1.4 \pm 3.1 \pm 2.5)\cdot 10^{-43}$ cm$^3$ & \\
\cline{2-4}
&
$\pi^- + (A,Z) \rightarrow \pi^- \pi^0 + (A,Z)$&
Study of chiral anomaly  $F(\gamma \rightarrow 3\pi)=12.9 \pm  
\newline \pm 0.9 
\pm 0.5$ GeV$^{-3}$& [29]\\
\hline
CERN & $\Lambda + (A,Z) \rightarrow \Sigma^0 + (A,Z)~~ \langle P_{\Lambda}
\rangle=15$ GeV & $\Gamma[\Sigma^0 \rightarrow \Lambda + \gamma]
=7.6^{+2.0}_{-1.3}$ KeV & [30] \\ 
\hline 
Fermilab & $\Lambda + (A,Z) \rightarrow \Sigma^0 + (A,Z)~~ \langle P_{\Lambda}
\rangle=200$ GeV & $\Gamma[\Sigma^0 \rightarrow \Lambda + \gamma]
=8.6 \pm 0.6 \pm 0.8 $ KeV & [31] \\ \hline 
\end{tabular}
}
\end{table}


\newpage

\begin{table}
\vspace*{-2cm}
\small
\caption{Study of the coherent
reaction $\pi^- + (A,Z) \to (\pi^- \pi^- \pi^+)
+(A,Z)$ 
and the determination of radiative with $\Gamma[a_2(1320)^- \to \pi^- +\gamma]$.}
\label{tab-1}
\rotate{
\begin{tabular}{|p{10mm}|p{28mm}|p{35mm}|p{25mm}||p{35mm}|p{35mm}|p{30mm}|}
\hline
Target & Number of events \newline $\pi^- + (A,Z) \to \newline 
\to (3\pi)^- + (A,Z)$
 & Number of $a_2(1320)$ events produced in the Coulomb process & 
$\Gamma[a_2(1320)^- \to 
\newline \to \pi^- +\gamma]$ (keV) & Average value \newline
$\Gamma[a_2(1320)^- \to \pi^- +\gamma]$ \newline (keV) &
$\Gamma[a_2(1320)^- \to \pi^- +\gamma]$ \newline from [21] (keV) &
Theoretical predictions for \newline $\Gamma[a_2(1320)^- \to 
\newline \to \pi^- + \gamma]$ (keV)\\
\hline
Carbon & $2760523$ & $1587 \pm 480$ & $289 \pm 87$ & & & 348 [37]\\ 
\cline{1-4}
Copper & $ 1997972 $& $ 5170 \pm 590 $& $ 248 \pm 27 $ & 
$225\pm 25({\rm stat.}) \pm  $
 &$ 295 \pm 60$&
\\ \cline{1-4}
Lead & 549092 & $2945 \pm 400$& $198\pm 27$ &
$\pm 45({\rm syst.}) = 225 \pm 51$& &
235[38]\\
\hline 
\end{tabular}
}
\end{table}

\begin{figure}[h]
\psfig{file=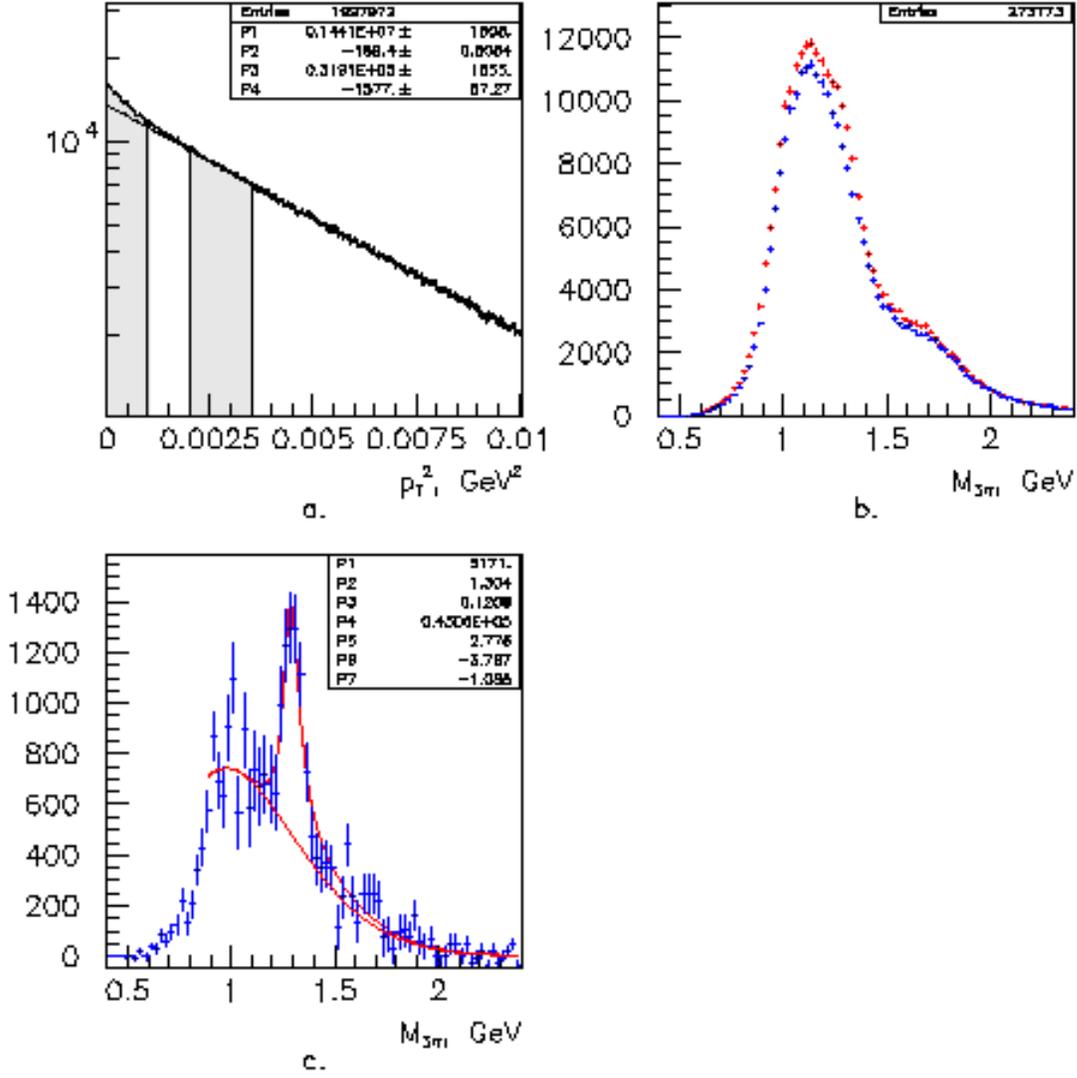,width=\textwidth}

\vspace*{-0.2cm}\parbox[c]{16cm}{\caption{\footnotesize The data for the 
coherent reaction $\pi^- + Cu \to [\pi^- \pi^- \pi^+]
+ Cu$ at $E_{\pi^-}=600$ GeV obtained at the SELEX setup.
a) $dN/dP^2_T$ distribution. The dash line shows the slope for
coherent diffraction process caused by the Pomeron exchange.
Region 1 ($P^2_T<0.001$ GeV$^2$) and region 2
($0.0015<P^2_T<0.0035$)  are used for the Coulomb production separation 
and for subtraction of the diffractive coherent background.
b) Effective mass spectra $M(3 \pi)$ for region 1 (red line)
and for region 2 (blue line).
c) Effective mass spectrum $M(3 \pi)$ for the Coulomb production
process, obtained by subtraction $M(3 \pi)_1 - a \cdot M(3 \pi)_2$
after a proper normalization ($a$ --- the normalization factor which was
chosen in  such a way that the number of diffractive background events in 
$N(3 \pi)_1$ and $a \cdot N(3 \pi)_2$ were identical).}}
\end{figure}

\end{document}